\documentclass{mem}
\usepackage{natbib}\usepackage{txfonts}\usepackage{balance}
\usepackage{graphicx}
\usepackage[a4paper]{hyperref}
\begin{document}
\def\teff{$T\rm_{eff }$}
\def\kms{$\mathrm {km s}^{-1}$}

\title{
Stellar evolution and the `O-rich AGB sequence'
}

   \subtitle{}

\author{
F.M. \,Jim\'{e}nez-Esteban\inst{1}
\and D. \,Engels\inst{1}
\and P. \,Garc\'{\i}a-Lario\inst{2}
          }

  \offprints{F.M. \,Jim\'{e}nez-Esteban\\ \email{steb313@hs.uni-hamburg.de}}

\institute{
Hamburger Sternwarte, Gojenbergsweg 112, D-21029 Hamburg, Germany.
        \and ISO Data Centre / European Space Astronomy Center, Research and Scientific Support Department of ESA, Villafranca del Castillo, Apartado de Correos 50727, E-28080 Madrid, Spain.
}

\authorrunning{Jim\'{e}nez-Esteban}

\titlerunning{Stellar evolution and the `O-rich AGB sequence'}

\abstract{
The `O-rich AGB sequence' is a well defined area occupied by
oxygen-rich AGB stars (OH/IR) in the IRAS two-colour diagram [12]-[25]
vs [25]-[60] (Fig.\,\ref{iras2c}). This sequence is interpreted in terms of
evolutionary stage and/or progenitor mass with the aim to link
observed classes of AGB stars and Planetary Nebulae (PNe).
\keywords{Stars: OH/IR -- Stars: AGB and post-AGB -- Stars:
 circumstellar matter -- Stars: variable -- Stars: evolution --
 Infrared: stars}
}
\maketitle{}

\section{Introduction}

One of the most fundamental questions that still remains open in the
understanding of AGB stellar evolution is the interpretation of the
`O-rich AGB sequence'. It has been shown by several authors that this
sequence reflects the increase of optical thickness of the
circumstellar envelope as the star evolves along the AGB
(e.g. \citealt{Bedijn87,Volk88}).


\begin{figure}[h]
\resizebox{\hsize}{!}{\includegraphics[clip=true]{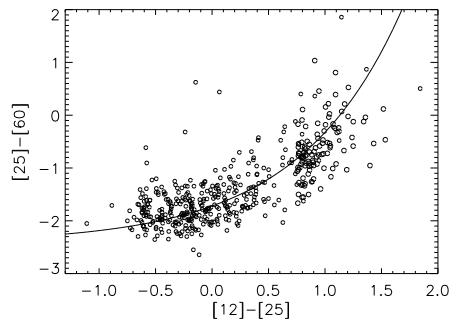}}
\caption{\footnotesize 
The position in the IRAS two-colour diagram of
OH/IR stars in the sample. The solid line is the
`O-rich AGB sequence' (see text), and the IRAS colours are defined as:
[12]$-$[25]\,=\,$-$2.5\,log$\frac{F_{\nu}(12)}{F_{\nu}(25)}$ and
[25]$-$[60]\,=\,$-$2.5\,log$\frac{F_{\nu}(25)}{F_{\nu}(60)}$.
}
\label{iras2c}
\end{figure}


However, this increase can be interpreted in two different ways: \\ 
i) {\bf evolutionary sequence}: every star would start the AGB phase
at the blue end of this sequence, and would later move while
increasing its mass-loss rate until reaching the red extreme at the
end of the AGB; \\ 
ii) {\bf mass sequence}: their different location would be just a
consequence of their different initial mass, which would determine the
mass-loss rate; \\
iii) A third interpretation was proposed, which is a combination of
the previous ones: all AGB stars would move towards redder colours as
they increase their mass-loss rate, but only the more massive stars
would be able to reach the reddest end of the sequence.

In two previous works \citep{Jimenez-Esteban05a,Jimenez-Esteban05b},
we have studied a very large sample of OH/IR stars with a very good
coverage of the whole `O-rich AGB sequence' (Fig.\,\ref{iras2c}). We
also have photometric information not only in the mid- and
far-infrared but also in the near-infrared where the bluer objects
have the maximum of their spectral energy distribution.

\section{Luminosity, distance and galactic height}

Bolometric fluxes were obtained by integrating the photometric data
available from the near infrared to the far infrared domain (2MASS,
own photometry, MSX and IRAS data), and extrapolating both toward
shorter and longer wavelengths. The typical uncertainty is
$\approx$\,40\%.

\begin{figure}[h]
\resizebox{\hsize}{!}{\includegraphics[clip=true]{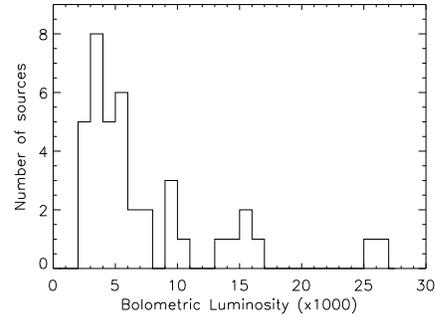}}
\caption{\footnotesize 
Absolute bolometric luminosity in thousands of {L$_{\odot}$} of the
OH/IR stars belonging to the Galactic Bulge population.
}
\label{LabsGB}
\end{figure}

We selected from our sample 41 extremely red OH/IR stars detected in
the direction of the Galactic Bulge, and assumed a common distance
(8\,kpc) to all them. The range of absolute luminosities obtained
(Fig.\,\ref{LabsGB}) is strongly peaked around 3\,500\,{L$_{\odot}$},
in agreement with those found by other authors using different OH/IR
star samples in the bulge with very different (bluer) colours
(e.g. \citealt{Wood98,Blommaert98}) or using samples located in
different parts of the Galaxy \citep{Knauer01}. Then we conclude
that the luminosity function may be similar throughout the Galaxy and
not very dependent on the colours of the stars selected. Our subsample
of OH/IR stars in the Galactic Bulge is dominated by relative low mass
stars, so, these stars can also reach the red part of the `O-rich AGB
sequence'.


We estimated the galactic height for the rest of OH/IR stars by simply
assuming a common and constant luminosity L$_{OH/IR}$ of
3\,500\,{L$_{\odot}$} for all them.

\section{Interpretation of the `O-rich AGB sequence'}

In order to have a good descriptor for the whole `O-rich AGB
sequence', we have used the parametrization introduced by the
following equations:

\begin{center}
$ \left. \begin{array}{l}
$[12]$-$[25]\,=\,0.912\,Ln\,$\lambda \\
\\
$[25]$-$[60]\,=\,$-$2.42\,+\,0.72\,$\lambda
\end{array} \right \} $
\end{center}

Each point of the `O-rich AGB sequence' can then be associated with a
given value of $\lambda$, and each star is then assigned a $\lambda$
value which corresponds to the nearest point in the `O-rich AGB
sequence'.


\begin{figure}[h]
\resizebox{\hsize}{!}{\includegraphics[clip=true]{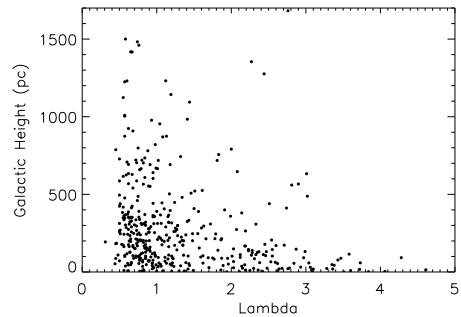}}
\caption{\footnotesize 
Galactic height distribution as a function of $\lambda$ derived
for all the sources in the sample, excluding the
Galactic Bulge OH/IR stars (L$_{OH/IR}$\,=\,3\,500\,{L$_{\odot}$} assumed).
}
\label{ZvsLamb}
\end{figure}


Fig.\,\ref{ZvsLamb} shows the galactic height in absolute value
derived as a function of the $\lambda$ parameter. We found a very
clear correlation, which implies that the redder part of the `O-rich
AGB sequence' must be populated mainly with objects of higher mass. In
addition, we found a similar correlation between {v$_{exp}$} and
$\lambda$. Note that assuming a different luminosity would just change
the scale of the y-axis.

The above results are only consistent with an evolutionary scenario in
which all OH/IR stars would start the AGB phase, independent of their
progenitor mass, in the bluer part of the `O-rich AGB sequence' and
then they would evolve toward redder colors, although only the more
massive stars would reach the very end of the sequence.

\begin{figure}[h]
\resizebox{\hsize}{!}{\includegraphics[clip=true]{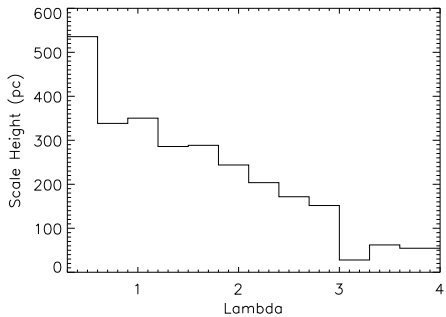}}
\caption{\footnotesize 
Galactic scale height distribution as a function of the
$\lambda$ parameter for all the sources in the sample,
excluding the Galactic Bulge.
}
\label{HvsLamb}
\end{figure}

In order to further characterise the stars populating the `O-rich AGB
sequence', we have divided the sequence in bins of $\lambda$ and
determined the galactic scale height H associated
(Fig.\,\ref{HvsLamb}). Then we have classified all the OH/IR stars in
our sample in 5 main groups.


The {\it `extremely blue subsample'} ($\lambda$\,$<$\,0.6;
H\,=\,536\,pc) is dominated by OH/IR stars that can be identified as
the result of the evolution of low mass optically bright short period
($\la$\,300\,days) Miras \citep{Jura94}. These OH/IR stars should be the
progenitors of so-called Type\,III PNe \citep{Maciel92}.

The {\it `blue subsample'} (0.6\,$<$\,$\lambda$\,$\le$\,1.2;
H\,=\,344\,pc) is mainly formed by OH/IR stars identified as the
result of the evolution of intermediate period ($\approx$\,300\,days)
Miras \citep{Wood77}. These OH/IR stars should be the progenitor
of low mass (O-rich) Type\,II PNe \citep{Maciel92}.

The {\it `transition subsample'} (1.2\,$<$\,$\lambda$\,$\le$\,1.8;
H\,=\,287\,pc) populates the region in which OH/IR stars become
optically thick. These OH/IR stars are the result of the evolution of
intermediate period (300\,-\,500\,days) Miras \citep{Jura93}, will
probably transform into C-rich AGB stars \citep{Groenewegen92}, and
must be the progenitors of high mass (C-rich) Type\,II PNe
\citep{Maciel92}.

The {\it `red subsample'} (1.8\,$<$\,$\lambda$\,$\le$\,3.0;
H\,=\,193\,pc) is formed by OH/IR stars with optically thick
circumstellar envelope. These stars are probaly undergoing Hot Botton
Burning proces, and should be the progenitors of Type\,I PNe
\citep{Maciel92}.

The {\it `extremely red subsample'} (3.0\,$<$\,$\lambda$;
H\,=\,48\,pc) is identified as the group containing the most massive
OH/IR stars in our sample. They should be the precursors of the
so-called OHPNe, heavily obscured OH/IR stars with radiocontinuum
emission that have been proposed to be infrared PNe.

\bibliographystyle{aa}

\end{document}